\ifx\documentclass\undefined
\documentstyle[12pt]{article}
\else
\documentclass[12pt]{article}
\fi

\makeatletter
\renewcommand{\theequation}{\thesection.\arabic{equation}}
\@addtoreset{equation}{section}
\def\eqnarray{%
\stepcounter{equation}%
\let\@currentlabel=\theequation
\global\@eqnswtrue
\global\@eqcnt\z@
\tabskip\@centering
\let\\=\@eqncr
$$\halign to \displaywidth\bgroup\@eqnsel\hskip\@centering
$\displaystyle\tabskip\z@{##}$&\global\@eqcnt\@ne
\hfil$\displaystyle{{}##{}}$\hfil
&\global\@eqcnt\tw@$\displaystyle\tabskip\z@{##}$\hfil
\tabskip\@centering&\llap{##}\tabskip\z@\cr}
\makeatother

\def\bbbz{{\mathchoice {\hbox{$\sf\textstyle Z\kern-0.4em Z$}}
{\hbox{$\sf\textstyle Z\kern-0.4em Z$}}
{\hbox{$\sf\scriptstyle Z\kern-0.3em Z$}}
{\hbox{$\sf\scriptscriptstyle Z\kern-0.2em Z$}}}}
\def\bbbq{{\mathchoice {\setbox0=\hbox{$\displaystyle\rm Q$}\hbox{\raise
0.15\ht0\hbox to0pt{\kern0.4\wd0\vrule height0.8\ht0\hss}\box0}}
{\setbox0=\hbox{$\textstyle\rm Q$}\hbox{\raise
0.15\ht0\hbox to0pt{\kern0.4\wd0\vrule height0.8\ht0\hss}\box0}}
{\setbox0=\hbox{$\scriptstyle\rm Q$}\hbox{\raise
0.15\ht0\hbox to0pt{\kern0.4\wd0\vrule height0.7\ht0\hss}\box0}}
{\setbox0=\hbox{$\scriptscriptstyle\rm Q$}\hbox{\raise
0.15\ht0\hbox to0pt{\kern0.4\wd0\vrule height0.7\ht0\hss}\box0}}}}
\def\bbbc{{\mathchoice {\setbox0=\hbox{$\displaystyle \rm C$}\hbox{\raise
0.06\ht0\hbox to0pt{\kern0.4\wd0\vrule height0.9\ht0\hss}\box0}}
{\setbox0=\hbox{$\textstyle\rm C$}\hbox{\raise
0.06\ht0\hbox to0pt{\kern0.4\wd0\vrule height0.9\ht0\hss}\box0}}
{\setbox0=\hbox{$\scriptstyle\rm C$}\hbox{\raise
0.06\ht0\hbox to0pt{\kern0.4\wd0\vrule height0.8\ht0\hss}\box0}}
{\setbox0=\hbox{$\scriptscriptstyle\rm C$}\hbox{\raise
0.06\ht0\hbox to0pt{\kern0.4\wd0\vrule height0.8\ht0\hss}\box0}}}}
\makeatletter
  \renewcommand{\theequation}{%
 \thesection.\arabic{equation}}
  \@addtoreset{equation}{section}
\makeatother
\newtheorem{theorem}{Theorem}[section]
\newtheorem{lemma}[theorem]{Lemma}
\newtheorem{corollary}[theorem]{Corollary}
\newtheorem{proposition}[theorem]{Proposition}
\newtheorem{remark}[theorem]{Remark}

\newsavebox{\toy}
\savebox{\toy}{\framebox[0.65em]{\rule{0cm}{1ex}}}
\newcommand{\QED}{\usebox{\toy}}
\def\nlni{\par\ifvmode\removelastskip\fi\vskip\baselineskip\noindent}
\newenvironment{proof}{\nlni\begingroup\it Proof.\rm}{
\endgroup\vskip\baselineskip}

\begin{document}
\setlength{\baselineskip}{15pt}
\title{
Distribution of localization centers in some discrete random systems
}
\author{Fumihiko Nakano
\thanks{Faculty of Science, 
Department of Mathematics and Information Science,
Kochi University,
2-5-1, Akebonomachi, Kochi, 780-8520, Japan.
e-mail : 
nakano@math.kochi-u.ac.jp}}
\date{}
\maketitle
\begin{abstract}
As a supplement of our previous work \cite{Killip-Nakano}, 
we consider the localized region of the random Schr\"odinger operators on 
$l^2( {\bf Z}^d )$
and study the point process composed of their eigenvalues and corresponding localization centers. 
For the Anderson model 
we show that, this point process in the natural scaling limit converges in distribution to the Poisson process on the product space of energy and space.
In other models with suitable Wegner-type bounds, 
we can at least show that limiting point processes are infinitely divisible. 
\end{abstract}


\section{Introduction}
The typical model 
we consider is the so-called Anderson model given below. 
\[
(H_{\omega}\varphi)(x) 
=
\sum_{|x-y|=1} \varphi(y) + \lambda V_{\omega}(x) \varphi(x), 
\quad
\varphi \in l^2 ({\bf Z}^d)
\]
where 
$\lambda > 0$
is the coupling constant and 
$\{ V_{\omega} (x) \}_{x \in {\bf Z}^d}$
are the independent, identically distributed random variables on a probability space
$(\Omega, {\cal F}, {\bf P})$.
The following facts are well-known. 

(1)
(the spectrum of 
$H$)
the spectrum of 
$H_{\omega}$
is deterministic almost surely
\[
\sigma (H_{\omega}) = \Sigma :=[-2d, 2d] + \lambda\mbox{ supp } d \nu, 
\quad
a.s.
\]
where 
$\nu$
is the distribution of 
$V_{\omega}(0)$
\cite{Kunz-Souillard}.

(2)(Anderson localization)
There is an open interval 
$I \subset \Sigma$
such that with probability one, 
the spectrum of 
$H_{\omega}$
on 
$I$
is pure point with exponentially decaying eigenfunctions. 
$I$
can be taken 
(i)
$I = \Sigma$
if 
$\lambda$
is large enough, 
(ii)
on the band edges,
and
(iii)
away from the spectrum of the free Laplacian if
$\lambda$
sufficiently small
(e.g., \cite{FS, vonDK, Ai, AM}).
\\
Recently, 
some relations between the eigenvalues and the corresponding localization centers are discussed\cite{repulsion}. 
It roughly implies,\\
(1)
If 
$|E - E_0| \simeq L^{-d}$ ($E_0 \in I$), 
the localization center
$x(E)$
corresponding to the energy 
$E$
satisfies
$|x(E)| \ge L$. 
Hence
the distribution of the localization centers are ``thin" in space\footnote{
This result follows easily from the upper bound on the density of states.
So in the Lifschitz tail region, we have 
$| x(E) | \ge (const.) e^{(const.) L^{\frac {d^2}{2}}}$
if
$| E - E_0 | \le L$}.\\
(2)
If
$| E - E' | \simeq L^{-2d}$, 
the localization centers 
$x(E), x(E')$
corresponding to the energies
$E,E'$
satisfies
$|x(E) - x(E')| \ge L$.
Hence 
the localization centers are repulsive if the energies get closer.

On the other hand, in
\cite{Killip-Nakano}, 
they study the ``natural scaling limit" of the  random measure in 
${\bf R}^{d+1}$
(the product of energy and space)
composed of the eigenvalues and eigenfunctions.
The result there roughly implies that
their distribution with eigenvalues in the order of 
$L^{-d}$
from the reference energy 
$E_0$, 
and with eigenfunctions in the order of 
$L$
from the origin,  obey the Poisson law on 
${\bf R}^{d+1}$. 
This work
can also be regarded as an extension of the work by Minami
\cite{Minami}
who showed that the point process on 
${\bf R}$
composed of the eigenvalues of 
$H$
in the finite volume approximation converges to the Poisson process on 
${\bf R}$. 
To summarize, 
\cite{repulsion, Killip-Nakano}
imply that the eigenfunctions whose energies are in the order of 
$L^{-d}$
are non-repulsive while those in the order of 
$L^{-2d}$
are repulsive, which are consistent with Minami's result
\cite{Minami}.

The aim 
of this paper is to supplement \cite{Killip-Nakano} from a technical point of view : 
(i)
to study the distribution of the localization centers which is technically different from what is done in \cite{Killip-Nakano}, 
and 
(ii)
to study what can be said for those models in which Minami's estimate and the fractional moment bound, which are the main tool in \cite{Killip-Nakano}, are currently not known to hold. 

We set some notations. 
\\

\noindent
{\bf Notation : }\\
(1)
For 
$x = (x_1, x_2, \cdots, x_d) \in {\bf Z}^d$, 
let 
$| x | = \sum_{j=1}^d | x_j |$.
$\Lambda_L(x) := \{ y \in {\bf Z}^d : 
|x - y | \le \frac L2 \}$
is the finite box in 
${\bf Z}^d$
with length 
$L$
centered at  
$x \in {\bf Z}^d$. 
$| \Lambda | := \sharp \Lambda$
is the number of sites in the box
$\Lambda$
and 
$\chi_{\Lambda}$
is the characteristic function of 
$\Lambda$.
\\
(2)
For a box
$\Lambda$, 
let
\begin{eqnarray*}
\tilde{\partial} \Lambda & := &
\left\{
\langle y, y' \rangle \in \Lambda \times \Lambda^c : 
|y - y'| =1
\right\}
\\
\partial \Lambda & := &
\left\{
y \in \Lambda : 
\langle y, y' \rangle \in \tilde{\partial} \Lambda
\mbox{ for some }
y' \in \Lambda^c
\right\}
\end{eqnarray*}
be two notions of the boundary of 
$\Lambda$. \\
(3)
For a box
$\Lambda(\subset {\bf Z}^d)$, 
$H_{\Lambda} := H |_{\Lambda}$
is the restriction of 
$H$
on 
$\Lambda$.
We consider both 
Dirichlet b.c. and periodic b.c. depending on cases, 
to be specified in which they are defined. 
For
$E \notin \sigma (H_{\Lambda})$, 
$G_{\Lambda}(E; x,y) = \langle \delta_x, (H_{\Lambda}-E)^{-1} \delta_y \rangle_{l^2(\Lambda)}$
is the Green function of 
$H_{\Lambda}$.
$\delta_x \in l^2({\bf Z}^d)$
is defined by
$\delta_x (y) = 1(y=x), =0 (y \ne x)$
and 
$\langle \cdot, \cdot \rangle_{l^2(\Lambda)}$
is the inner-product on 
$l^2(\Lambda)$.
\\
(4)
Let
$\gamma> 0, E \in {\bf R}$.
We say that the box
$\Lambda_L(x)$
is
$(\gamma, E)$-regular
iff
$E \notin \sigma(H_{\Lambda_L(x)})$
and the following estimate holds
\footnote{We adopt
this definition to treat Lemma \ref{Step 5} and Proposition \ref{infinite divisibility2}.}
\[
\sup_{\epsilon > 0}
| G_{\Lambda_L(x)} (E+ i \epsilon ; x, y) | 
\le e^{- \gamma \frac L2},
\quad
\forall y \in \partial \Lambda_L(x).
\]
In 
(\ref{MSA}), (\ref{eventk})
and
(\ref{Omegakdash}), 
we will consider this condition for 
$H_{\Lambda_L (x)}$
with Dirichlet b.c., while in 
(\ref{Gk}), (\ref{MSA for any boxes})
with periodic b.c.\footnote{with a slight change of argument, it is possible to set periodic b.c. only in this condition.}\\
(5)
For 
$\phi \in l^2({\bf Z}^d)$, $\phi \ne 0$, 
we define the set 
$X(\phi)$
of its localization centers by 
\[
X(\phi) := \left\{
x \in {\bf Z}^d : 
| \phi (x) | = \max_{y \in {\bf Z}^d} | \phi (y) |
\right\}
\]
This definition is due to 
\cite{BG}.
Since
$\phi \in l^2({\bf Z}^d)$, 
$X(\phi)$
is a finite set. 
To be free from ambiguities, we choose
$x(\phi) \in X(\phi)$
according to a certain order on
${\bf Z}^d$. 
For a box
$\Lambda$, 
we say 
$\phi$
is localized in 
$\Lambda$
iff 
$x(\phi) \in \Lambda$.
If 
$\{ E_j \}_j$, $\{ \phi_j \}_j$
are the enumerations of the eigenvalues and eigenfunctions of 
$H$
counting multiplicities, we set 
$X(E_j) := X(\phi_j), x(E_j) := x(\phi_j)$
and we say 
$E_j$
is localized in 
$\Lambda$
iff
$x(E_j) \in \Lambda$.
If an eigenvalue 
is degenerated, we adopt any but fixed selection procedure 
of choosing eigenfunctions so that the quantities in concern ($\bar{\xi}_k, \xi_k$
defined later) are measurable. 
\\
(6)
For a Hamiltonian 
$H$, 
an interval
$J (\subset {\bf R})$ 
and a box
$B( \subset {\bf Z}^d)$, 
we set 
\begin{eqnarray*}
{\cal E}(H, J)
& := &
\{
\mbox{ eigenvalues of $H$ in $J$}
\}\\
{\cal E}(H, J, B)
& := &
\{
\mbox{ eigenvalues of $H$ in $J$ localized in $B$ }
\}\\
{\cal E}f(H, J)
& := &
\{
\mbox{ normalized eigenfunctions of $H$ in $J$}
\}\\
{\cal E}f(H, J, B)
& := &
\{
\mbox{  normalized eigenfunctions of $H$ in $J$ localized in $B$ }
\}\\
N(H, J)
& := &
\sharp
{\cal E}(H, J)
\quad
\mbox{(counting multiplicity)}
\\
N(H, J, B)
& := &
\sharp
{\cal E}(H, J, B)
\end{eqnarray*}
(7)
While
our results(Theorem \ref{uniform distribution} and \ref{main theorem}) adopt 
$x(\phi)$
as a definition of localization center, 
a more natural definition of that may be
\[
\langle x \rangle_{\phi} := 
\sum_{y \in {\bf Z}^d} y | \phi (y) |^2
\left(
\sum_{y \in {\bf Z}^d}  | \phi (y) |^2
\right)^{-1}, 
\quad
\phi \in l^2({\bf Z}^d), 
\;
\phi \ne 0.
\]
However, 
it is easy to see (Lemma \ref{two notions}) that those theorems are also valid if we adopt 
$\langle x \rangle_{\phi}$ instead.\\
(8)
For a 
$n$-dimensional measurable set 
$A(\subset {\bf R}^n)$, 
we denote by 
$|A|$
its Lebesgue measure. 
For
$a \in {\bf R}$
and 
$r > 0$, 
$I(a, r) := \{ x \in  {\bf R} : | x - a | < r \}$
is the open interval centered at 
$a$
with radius 
$r$. 
\\
(9)
Set 
$K = [0, 1]^d$
and let
$\pi_e$ 
and 
$\pi_s$
be the canonical projections on 
${\bf R} \times K$
onto 
${\bf R}$
and  
$K$
respectively : 
$\pi_e(E,x) = E$, $\pi_s(E, x) = x$
for
$(E, x) \in {\bf R} \times K$.
\\
(10)
We set 
\[
\xi (f) = \int_{{\bf R}\times K} f(x) \xi (dx)
\]
for a Radon measure
$\xi$
and a bounded measurable function 
$f$
on 
${\bf R} \times K$. 
Even if
$f$
is a function on 
${\bf R}$, 
we write
$\xi (f)$
instead of 
$\xi(f 1_K)$
for simplicity. 
For a sequence
$\{ \xi_k \}_k$
of Radon measures, 
$\xi_n \stackrel{v}{\to} \xi$
means 
$\xi_n$
converges vaguely to 
$\xi$ : 
$\xi_n(f) \stackrel{n \to \infty}{\to} \xi(f)$
for any
$f \in C_c({\bf R} \times K)$.\\

We consider the following two assumptions. \\

\noindent
{\bf Assumption A}
\\
{\it (1)
(Initial length scale estimate)
Let 
$I (\subset \Sigma)$
be an open interval where the initial length scale estimate of the multiscale analysis holds : 
we can find
$\gamma > 0$
and 
$p > 6d$
such that for sufficiently large
$L_0$
we have
\[
{\bf P}\left(
\mbox{ For any } E \in I, 
\Lambda_{L_0}(0)
\mbox{ is $(\gamma, E)$-regular }
\right)
\ge 1 - L_0^{-p}.
\]
where
$H_{\Lambda_{L_0}(0)}$
has Dirichlet b.c.\\
(2)(Wegner's estimate)
We can find a positive constant
$C_W$
such that for any interval 
$J (\subset I)$
and any box
$\Lambda$, 
\begin{equation}
{\bf E}[ N(H_{\Lambda}, J) ] \le C_W | \Lambda | | J |.
\label{Wegner's estimate}
\end{equation}
In (2), 
we require that both 
$H_{\Lambda}$'s
with Dirichlet b.c. and those with periodic b.c. satisfy 
(\ref{Wegner's estimate}).
}

Assumption A
is known to hold,  for instance, 
(1)
for the Anderson model when the distribution of the random potential 
$\nu$
has the bounded density 
$\rho$, 
with the allowed location of 
$I$
mentioned at the beginning of this section
\cite{Wegner, vonDK}, 
(2)
for the Schr\"odinger operators with off-diagonal disorder
\cite{Faris}, and
(3)
for the Schr\"odinger operators on 
$l^2({\bf Z}^2)$
with random magnetic fluxes\cite{KNNN}
(in (2), (3), 
$I$
is on the edge of $\Sigma$).
We need 
$p > 6d$
to eliminate the contributions from the negligible events, in the proof of 
Proposition \ref{infinite divisibility}.

Pick
$\alpha$
with 
$1 < \alpha < \alpha_0 := \frac {2p}{p+2d}(<2)$, 
and set
\[
L_{k+1} = L_k^{\alpha}, 
\quad
k = 0, 1, \cdots
\]
For simplicity, we write
$\Lambda_k(x) = \Lambda_{L_k}(x)$.
By the multiscale analysis \cite{vonDK}, we have, for 
$k = 1, 2, \cdots$
and for any fixed disjoint boxes 
$\Lambda_{k}(x), \Lambda_{k}(y)$, 
\begin{eqnarray}
{\bf P}
\Biggl(
&&
\mbox{ For any 
$E \in I$, 
either
$\Lambda_k(x)$
or
$\Lambda_k(y)$
are 
$(\gamma, E)$-regular
}
\Biggr)
\nonumber
\\
&&
\qquad
\ge
1 - L_k^{-2p}.
\label{MSA}
\end{eqnarray}
where we take Dirichlet b.c. for 
$H_{\Lambda_k (x)}$, $H_{\Lambda_k(y)}$.
\\

\noindent
{\bf Assumption B} 
(Minami's estimate)\\
{\it 
We can find a positive constant 
$C_M$
such that for any finite box
$\Lambda$
and any interval 
$J (\subset I)$, 
\[
\sum_{k=2}^{\infty}
k(k-1)
{\bf P}
\left(
N(H_{\Lambda}, J) = k
\right)
\le
C_M | \Lambda |^2 | J |^2
\]
where
$H_{\Lambda}$
has periodic b.c.
}
Assumption B is known to be true for the Anderson model and for any interval 
$J (\subset {\bf R})$ 
when $\nu$ has the bounded density\cite{Minami}.

The integrated density of states
$N(E)$ 
of 
$H$
is defined by
\begin{equation}
N(E) :=
\lim_{| \Lambda | \to \infty}
\frac {1}{| \Lambda |}
N(H_{\Lambda}, (-\infty, E]).
\label{IDS}
\end{equation}
It is known that, 
with probability one, 
this limit exists for any 
$E \in {\bf R}$ 
and continuous
\cite{CL}
so that its derivative
$n(E)$
finitely exists a.e.
which is called the density of states.

Let 
$M({\bf R}^n)$ (resp. $M_p ({\bf R}^n)$)
be the set of Radon measures (resp. integer-valued Radon measures) on 
${\bf R}^n$
which is regarded as a metric space under the 
vague topology.
Random measure (resp. point process) on 
${\bf R}^n$
is defined to be a measurable mapping from 
$(\Omega, {\cal F}, {\bf P})$
to
$M ({\bf R}^n)$
(resp. $M_p ({\bf R}^n)$).
We say that a sequence 
$\{ \xi_k \}_k$
of random measures converges in distribution to a random measure
$\xi$
and write
$\xi_k \stackrel{d}{\to} \xi$
iff the distribution of 
$\xi_k$
converges weakly to that of 
$\xi$.
We state our results below. \\

\noindent
(1)
{\bf Uniform distribution of localization centers : }
We first consider localization centers corresponding to all eigenvalues in $I$.
Let 
$H_k = H |_{\Lambda_k}$
be the restriction of 
$H$
on 
$\Lambda_k = \{1, 2, \cdots, L_k\}^d$
with periodic boundary condition.
The choice 
of this particular boundary condition is to be free from the boundary effect which should be purely technical. 
Writing 
$\{ F_j (\Lambda_k) \}_j : = {\cal E}(H_k, I)$, 
we define a random measure 
$\bar{\xi}_k$
on 
$I \times K$
by
\[
\bar{\xi}_k := \frac {1}{| \Lambda_k |}
\sum_j \delta_{\bar{X_j}}, 
\quad
\bar{X_j} := \left(
F_j (\Lambda_k), L_{k}^{-1} x(F_j(\Lambda_k))
\right)
\in I \times K.
\]
\begin{theorem}
\label{uniform distribution}
Assume Assumption A(1) with 
$p >2d$.
Then 
\label{uniform distribution}
\[
\bar{\xi}_k \stackrel{v}{\to} \nu \otimes dx,
\quad
a.s.
\]
\end{theorem}
Theorem \ref{uniform distribution}
implies that the localization centers are uniformly distributed in the macroscopic scale. 
Although 
the proof is straightforward by using the existence of the density of states, we provide that in Section 7 for completeness. 
In 
\cite[Theorem 7.1]{dR}, 
same conclusion is derived for a special case 
(i.e., eq. 
(\ref{suffice})
in Section 7 is proved for
$I = \Sigma$
and
$J = {\bf R}, B = K$), 
and in
\cite{Killip-Nakano}, 
almost equivalent statement is derived for all energies. \\

\noindent
(2)
{\bf Local fluctuation : }
To see 
the local fluctuation near a reference energy 
$E_0 \in I$, 
let
$\{ E_j(\Lambda_k) \}_j := {\cal E}(H_k, {\bf R})$, 
$\{ x_j \}_j := \{ x(E_j(\Lambda_k)) \}_j$
and define a point process 
$\xi_k$
on 
${\bf R} \times K$
as follows. 
\[
\xi_k = \sum_{j=1}^{| \Lambda_k |} \delta_{X_j},
\quad
X_j = (| \Lambda_k | (E_j(\Lambda_k) - E_0), L_k^{-1} x_j)
\in {\bf R} \times K.
\]
This scaling 
is the same as that in 
\cite{Minami, Killip-Nakano} : 
the energies are supposed to accumulate in the order of 
$L^{-d}$
around
$E_0$
for large 
$L$ 
if 
$n(E_0) < \infty$, 
and if 
$|E - E_0| \simeq L^{-d}$, 
we expect
$| x(E) | \simeq L$
\cite[Theorem 1.1]{repulsion}. 
The main theorem of this paper is 
\begin{theorem}
%
Assume
Assumptions A, B.
If 
$n(E_0) < \infty$, 
then
$\xi_k \stackrel{d}{\to} \zeta_{P, {\bf R} \times K}$
as 
$k \to \infty$
where 
$\zeta_{P, {\bf R} \times K}$
is the Poisson process on 
${\bf R} \times K$
with its intensity measure
$n(E_0) dE \times dx$.
\label{main theorem}
\end{theorem}
If we do not assume
Assumption B, 
we can only prove that there exists convergent subsequence and its limiting point process is infinitely divisible with absolutely continuous intensity measure(Theorem \ref{partial result}). 
(we say the point process
$\xi$
is infinitely divisible
iff for any 
$n \in {\bf N}$, 
we can find i.i.d. array of point process
$\{ \xi_{nj} \}_{j=1}^n$
with 
$\xi\stackrel{d}{=}\sum_{j=1}^n \xi_{nj})$.
Similar conclusion is proved in 
\cite{Giere}
for one-dimensional random
Schr\"odinger operator on 
${\bf R}$.
The infinite divisibility of 
$\xi$
merely implies that 
$\xi$
is represented by the Poisson process on 
$M_p ({\bf R} \times K)$
whose intensity measure is given by its 
canonical measure \cite[Lemma 6.5, 6.6]{Kallenberg}.
We are unable to prove 
Theorem \ref{main theorem}
if we replace 
$H_k$
by 
$H$
itself (which is done in \cite{Killip-Nakano})
because some ``a priori" estimates are missing to prove Step 1 in the proof of Proposition \ref{infinite divisibility}, 
Lemma \ref{many eigenvalues}
and
Lemma \ref{many eigenvalues3}.

By ``projecting"
the result of Theorem \ref{main theorem} to the energy axis, we recover the result by Minami \cite{Minami} : 
the point process 
$\xi_k^{(ev)} = \sum_j 
\delta_{| \Lambda_k |(E_j(\Lambda_k) - E_0)}$
converges to the Poisson process on 
${\bf R}$.  
If we project it to the space axis, 
we have a result on the distribution of localization centers. 
To be precise, 
pick an interval 
$J (\subset {\bf R})$
and let
$\{ F_j (\Lambda_k, J) \}_{j \ge 1}
:=
{\cal E}(H_k, E_0 + L_k^{-d}J)$.
Define a point process on 
$K$
by
\[
\xi_{k}^{(loc)} = \xi_k(J \times \cdot)
=
\sum_{j \ge 1} \delta_{L_k^{-1} x_j(F_j (\Lambda_k, J))}.
\]
\begin{corollary}
Under the same assumption as in 
Theorem \ref{main theorem}, 
$\xi_{k}^{(loc)} \stackrel{d}{\to} \zeta_{P, K}$
as
$k \to \infty$
where 
$\zeta_{P, K}$
is the Poisson process on 
$K$
with intensity measure
$n(E_0) | J |  dx$.
\end{corollary}
In \cite{KLP}, 
they assume the Poisson distribution of localization centers and discuss the derivation of Mott's formula on the a.c. conductivity(rigorous derivation of Mott's formula is recently done by \cite{KLM}). 

The remaining sections
are organized as follows. 
In Section 2, 
we prove the infinite divisibility of the limiting point process (Theorem \ref{partial result})
which is one of the main condition to apply the Poisson convergence theorem \cite[Corollary 7.5]{Kallenberg} to our situation.
In order to do that, 
we decompose 
$\Lambda_k$
into disjoint boxes 
$\{ D_p \}_p$
of size 
$L_{k-1}$, 
and let 
$H_p = H |_{D_p}$
as is done in 
\cite{Minami}.
Since 
the eigenfunctions of 
$H_k$
corresponding to the eigenvalue
$E$
 in 
$I$
are exponentially localized, we can find a box
$D_p$
such that 
$H_p$
has eigenvalues near
$E$. 
By some perturbative argument, 
we can construct a one to one correspondence between the eigenvalues of 
$H_k$
and that of 
$\oplus_p H_p$, 
with probability close to  $1$. 
Therefore, 
$\xi_k$
is approximated by the sum 
$\eta_k = \sum_p \eta_{k, p}$
of the point process composed of the eigenvalues and localization centers of 
$H_p$. 
Wegner's estimate 
ensures that 
$\{ \eta_{k, p} \}_{k, p}$
is a null-array and relatively compact, so that 
$\{ \eta_k \}_k$ 
always has the convergent subsequence whose limiting point is infinitely divisible.

In Section 3, under Assumption A, B, 
we show that 
$\eta_k$
converges in distribution to the Poisson process, finishing the proof of 
Theorem \ref{main theorem}.  
By
Minami's estimate, 
$\eta_{k, p}$
has at most one atom in the corresponding region in 
${\bf R} \times K$
with the probability close to $1$. 
Hence the general Poisson convergence theorem 
\cite[Corollary 7.5]{Kallenberg}
gives the result. 
Since the mechanism 
of the convergence to the Poisson process is the same as in 
\cite{Minami}, 
Theorem \ref{main theorem}
can be regarded as the extension of that. 

To construct 
that one to one correspondence, we used the machinery developed in \cite{BG, KM}
which is reviewed in Section 4. 

For the random measure studied in 
\cite{Killip-Nakano}, 
we can show the infinite divisibility as Theorem \ref{partial result} under Assumption A, to be mentioned in Section 5. 

If we assume both  
Assumption A and B in the proof of 
Proposition \ref{infinite divisibility}, 
$H_{p}$
has at most one eigenvalues in the corresponding region with probability close to 
$1$, 
so that the correspondence between eigenvalues of 
$H_{k+1}$
and 
$H_{p}$ 
becomes bijective apart from negligible contributions, 
which is mentioned in Section 6. 
This observation allows us 
to construct explicitly the approximate eigenfunctions of 
$H$
by its finite volume operator\cite{approximation}. 
In Section 7, 
we prove Theorem \ref{uniform distribution}.
\section{Infinite Divisibility}
For simplicity, 
we consider 
$\xi_{k+1}$
instead of 
$\xi_k$. 
We first decompose
$\Lambda_{k+1}$
into disjoint cubes 
$D_p$
of size
$L_k$ : 
$\Lambda_{k+1} = \bigcup_{p=1}^{N_k} D_p$, 
$N_k = \left(
\frac {L_{k+1}}{L_k}
\right)^d (1 + o(1))$.
The contribution 
of $D_p$'s near the boundary of 
$\Lambda_{k+1}$
turns out to be negligible by 
Lemma \ref{pre priori estimate}.
We denote by 
$C_p$
the box obtained by eliminating the strip of width 
$L_{k-1}$
from the boundary of 
$D_p$ : 
\[
C_p :=
\{ x\in D_p : d(x, \partial D_p) \ge L_{k-1} \}.
\]
Let 
$H_{k, p} := H |_{D_p}$
with periodic boundary condition. 
We set the following event 
\begin{eqnarray}
\Omega_k &=& \Biggl\{ \omega \in \Omega : 
\mbox{For any} 
E \in I, 
\mbox{ 
either 
$\Lambda_{k-1}(x)$
or 
$\Lambda_{k-1}(y)$
are
$(\gamma, E)$-regular 
}
\nonumber
\\
&&
\mbox{for any disjoint pair of boxes } 
\Lambda_{k-1}(x), \Lambda_{k-1}(y)
\subset 
\Lambda_{k+1} \cup \bigcup_{p=1}^{N_k} D_p
\Biggr\}
\label{eventk}
\end{eqnarray}
($H_{\Lambda_{k-1}(x)}$, $H_{\Lambda_{k-1}(y)}$
have Dirichlet b.c.).
In (\ref{eventk}), 
we regard 
$\Lambda_{k+1}$
and
$D_p$'s
as torus 
(so that now
$\Lambda_{k+1} \ne \bigcup_{p=1}^{N_k} D_p$)
and consider all 
$\Lambda_{k-1}(x)$'s
contained in 
$\Lambda_{k+1}$
or some
$D_p$.
So, for
$x \in \Lambda_{k+1}$ 
close to 
$\partial \Lambda_{k+1}$
(or close to some 
$\partial D_p$), 
some portion of 
$\Lambda_{k-1}(x)$
may appear in the opposite side to 
$x$
of 
$\partial \Lambda_{k+1}$
(or
$\partial D_p$).
This peculiar definition of the event 
$\Omega_k$
is for the proof of  Lemma \ref{phi decays at boundary}. 
By (\ref{MSA}),  we have
\begin{equation}
{\bf P}(\Omega_k)
\ge
1 - (const.) L_{k-1}^{-2p} L_{k+1}^{2d}
=
1 - (const.) L_{k-1}^{-2p + 2 d \alpha^2}.
\label{Omegak}
\end{equation}
We define the point process by 
\begin{eqnarray*}
\eta_{k+1} &=& \sum_{p=1}^{N_k} 
\eta_{k+1, p},
\quad
\eta_{k+1, p}  = \sum_{j=1}^{| D_p |} \delta_{Y_{p, j}}
\\
Y_{p, j} &=& 
(| \Lambda_{k+1} | (E_j(D_p) - E_0), L_{k+1}^{-1} y_{p, j})
\end{eqnarray*}
where 
$\{ E_j (D_p) \}_j := {\cal E}(H_{k, p}, {\bf R})$, 
$y_{p, j} = x(E_j(D_p))$. 
As was explained in 
Introduction, we expect that 
$\xi_{k+1}$
can be approximated by 
$\sum_p \eta_{k+1, p}$
to be shown below.
\begin{proposition}
%
Under 
Assumption A, we have 
\label{infinite divisibility}
\[
{\bf E} \left[
| \xi_{k+1}(f) - \eta_{k+1}(f) |
\right]
\to 0, 
\quad
k \to \infty, 
\quad
f \in C_c({\bf R} \times K).
\]
\end{proposition}
By
Proposition \ref{infinite divisibility}, 
the Laplace transform 
$L_{\xi}(f) = {\bf E}[ e^{-\xi(f)} ]$
of
$\xi$
satisfies
$\lim_{k \to \infty}(
L_{\xi_{k+1}}(f) - L_{\eta_{k+1}}(f)
)=0$, 
for
f $\in C_c^+ ({\bf R}\times K)$.
Hence 
it suffices to show 
$
\eta_{k+1} \stackrel{d}{\to} \zeta_{P, {\bf R} \times K}
$
to prove 
Theorem \ref{main theorem}.
By choosing 
$f$
independent of the space variables, 
we obtain an alternative proof of 
\cite[Step 3]{Minami}.
Since we use 
the exponential decay of eigenfunctions instead of that of Green's function, 
this proof is mathematically indirect but physically direct. 
\begin{proof}
{\it Step 1} : 
We show the contribution by the event
$\Omega_k^c$
is negligible. 
In fact, since 
$| \xi_{k+1}(f) | \le \| f \|_{\infty} | \Lambda_{k+1} |$
and since
$p > 6d > \frac 32 d \alpha^2$, 
we have
\[
{\bf E}[ | \xi_{k+1}(f) | ; \Omega_k^c]
\le
\| f \|_{\infty} | \Lambda_{k+1} | L_{k-1}^{-2 p + 2 d \alpha^2}
=
(const.)
L_{k-1}^{-2p + 3 d \alpha^2} = o(1)
\]
by (\ref{Omegak})\footnote{
the equation
``$\cdots = o(1)$"
henceforth means 
``$\cdots = o(1)$ 
as $k \to \infty$".
}. 
${\bf E}[ \sum_p \eta_{k+1, p}(f) ; \Omega_k^c]$
can be estimated similarly. 
Therefore, it suffices to show 
\[
{\bf E} \left[
| \xi_{k+1}(f) - \eta_{k+1}(f) | ; \Omega_k
\right]
=o(1).
\]
\noindent
{\it Step 2} : 
We show the contribution by the atoms whose localization centers are in 
$\bigcup_p (D_p \setminus C_p)$
are negligible. 
We first decompose 
\begin{eqnarray*}
\xi_{k+1} &=& \xi_{k+1}^{(1)} + \xi_{k+1}^{(2)},
\quad
\xi_{k+1}^{(j)} = \sum_{p=1}^{N_k} \xi_{k+1, p}^{(j)}, 
\quad
j = 1,2, 
\\
\xi_{k+1, p}^{(1)}
&=&
\sum_{x_j \in C_p} \delta_{X_j}, 
\quad
\xi_{k+1, p}^{(2)}
=
\sum_{x_j \in D_p \setminus C_p} \delta_{X_j}.
\end{eqnarray*}
And we decompose 
$\eta_{k+1, p}$
similarly. 
In what follows, we take any 
$0 < \gamma ' < \gamma$
and let 
$k$
large enough with 
$k \ge k_2(\alpha, d, \gamma, \gamma') 
\vee
k_3(\alpha, d, \gamma, \gamma')$
where 
$k_2, k_3$
are defined in 
Lemma \ref{many eigenvalues}
and
\ref{many eigenvalues3}. 
For simplicity, set
\[
\epsilon_{k-1} := e^{- \gamma' L_{k-1}/2}.
\]

\noindent
{\bf Claim 1 }
\[
{\bf E} [ \xi_{k+1}^{(2)}(f) ; \Omega_k ] 
= o(1),
\quad
{\bf E} [ \eta_{k+1}^{(2)}(f) ; \Omega_k ] 
= o(1).
\]\\
{\it Proof of Claim 1 } 
Let
\[
S_p = \{ x \in \Lambda_{k+1} : 
d(x, \partial (D_p \setminus C_p)) \le L_{k-1} \}, 
\quad
H'_{k,p} := H_{k+1} |_{S_p}
\]
(with Dirichlet boundary condition).
Pick 
$a > 0$
with 
$\pi_e (\mbox{ supp } f) \subset [-a,a]$
and set 
\[
J_{k+1} := \left[
E_0 - \frac {a}{| \Lambda_{k+1} |}, 
E_0 + \frac {a}{| \Lambda_{k+1} |}
\right]
=
I\left(
E_0, \frac {a}{| \Lambda_{k+1} |}
\right).
\]
By Lemma \ref{many eigenvalues}(2) and Assumption A(2)(Wegner's estimate), 
we have
\begin{eqnarray*}
{\bf E}[\xi_{k+1, p}^{(2)} (f) ; \Omega_k ]
& \le &
\| f \|_{\infty}
{\bf E}[
N(H_{k+1}, J_{k+1}, D_p \setminus C_p) ]
\\
& \le &
\| f \|_{\infty}
{\bf E}[
N(H'_{k, p}, J_{k+1}+ I(0, \epsilon_{k-1})) ]
\\
& \le &
(const.)
\| f \|_{\infty} C_W
| D_p \setminus C_p | 
\cdot
\frac {2a}{| \Lambda_{k+1} |}.
\end{eqnarray*}
Using the inequality
$| D_p \setminus C_p | \le (const.) L_{k-1} L_k^{d-1}$
and then taking sum w.r.t. 
$p$
gives
\begin{eqnarray*}
{\bf E}[\xi_{k+1}^{(2)}(f); \Omega_k]
\le
(const.) \frac {L_{k-1}}{L_k} = o(1).
\end{eqnarray*}
To estimate 
$\eta_{k+1}^{(2)}$, 
we set
\[
T_p = 
\{ x \in D_p : d(x, \partial ( D_p \setminus C_p )) \le L_{k-1} \}, 
\quad
H_{k,p}'' := H_{k, p} |_{T_p}.
\]
Then
the same argument as above with
Lemma \ref{many eigenvalues}(3)
gives
${\bf E}[\eta_{k+1}^{(2)}(f); \Omega_k]
\le
(const.) \frac {L_{k-1}}{L_k} = o(1)$
and thus proves Claim 1.
\QED
\\

\noindent
Therefore, 
it suffices to show 
\[
{\bf E} \left[
| \xi_{k+1}^{(1)}(f) - \eta_{k+1} (f) | ; \Omega_k
\right]
= o(1).
\]
The equation
${\bf E} [ \eta_{k+1}^{(2)}(f) ; \Omega_k ] 
= o(1)$
in Claim 1 will be used in Step 3 below. \\

\noindent
{\it Step 3} : 
We first show the following claim.\\

\noindent
{\bf Claim 2 }
Let 
$J \subset I$
be an interval.
If 
$\omega \in \Omega_k$, 
we have
\begin{eqnarray*}
&&\sum_p
N(H_{k, p}, J +  I(0, \epsilon_{k-1}))
\\
&\le&
\sum_p
N(H_{k+1}, J, C_p)
+
\sum_p
N(H_{k+1}, (J +  I(0, 2\epsilon_{k-1}))\setminus J, C_p)
\\
&&\qquad
+
\sum_p
N(H'_{k, p}, J +  I(0, 3\epsilon_{k-1}))
+
\sum_p
N(H''_{k, p}, J +  I(0, 2\epsilon_{k-1})).
\end{eqnarray*}
{\it Proof of Claim 2 }
We decompose
\begin{eqnarray*}
N(H_{k, p}, J +  I(0, \epsilon_{k-1}))
&=&
N(H_{k, p}, J +  I(0, \epsilon_{k-1}), C_p)
+
N(H_{k, p}, J +  I(0, \epsilon_{k-1}), D_p \setminus C_p)
\\
&=:& I_p + II_p.
\end{eqnarray*}
By
Lemma \ref{many eigenvalues}(3), 
\[
II_p \le N(H''_{k, p}, J + I(0, 2\epsilon_{k-1}))
\]
and by 
Lemma \ref{many eigenvalues}(2)
and 
Lemma \ref{many eigenvalues3}, 
\begin{eqnarray*}
\sum_p I_p
& \le &
N(H_{k+1}, J +  I(0, 2\epsilon_{k-1}))
\\
&=&
\sum_p
N(H_{k+1}, J + I(0, 2\epsilon_{k-1}), C_p)
+
\sum_p
N(H_{k+1}, J +  I(0, 2\epsilon_{k-1}), D_p \setminus C_p)
\\
& \le &
\sum_p 
N(H_{k+1}, J +  I(0, 2\epsilon_{k-1}), C_p)
+
\sum_p
N(H'_{k, p}, J +  I(0, 3\epsilon_{k-1}))
\end{eqnarray*}
which shows Claim 2. 
\QED\\

For any 
$p = 1, 2, \cdots, N_k$, 
let
$\{ E_{p, j} \}_j
=
{\cal E}(H_{k+1}, J_{k+1}, C_p)$
and write
$\bigcup_j
I(E_{p, j}, \epsilon_{k-1})$
as the disjoint union of open intervals : 
\[
\bigcup_j
I(E_{p, j}, \epsilon_{k-1}) 
=
\bigcup_i I_i.
\]
If
$
a_1^{(i)} < a_2^{(i)} < \cdots < a_{N_i}^{(i)}
$
are the elements of 
${\cal E}(H_{k+1}, I_i, C_p)$, then
\[
I_i
=
I'_i + I(0, \epsilon_{k-1}), 
\quad
I'_i
:=
(a_1^{(i)}, a_{N_i}^{(i)}).
\]
Letting 
$J = I'_i$ 
in Lemma \ref{many eigenvalues}(1), 
we have
\[
N(H_{k+1}, I_i, C_p)
\le
N(H_{k, p}, I_i)
\]
and hence we have an one to one correspondence
$\Phi$ 
from
$\bigcup_i {\cal E}(H_{k+1}, I_i, C_p) =: \{ E_{j, p} \}_j$
to
$\bigcup_i {\cal E}(H_{k, p}, I_i)$.
Since
$\mbox{ diam } (I_i) \le L_k^d \epsilon_{k-1}$, 
they satisfy
\[
| E_{j, p} - \Phi(E_{j, p})| \le L_k^d  \epsilon_{k-1}.
\]
On the other hand, 
by letting 
$J = J_{k+1}$
in Lemma \ref{many eigenvalues}(1) and Claim 2 we see that, 
the number of elements of 
$\bigcup_p 
{\cal E}(H_{k, p}, J_{k+1} + I(0, \epsilon_{k-1}))$
which do not lie in the range of 
$\Phi$
is less than 
\begin{eqnarray*}
&&\sum_p 
N(H_{k+1}, (J_{k+1} +  I(0, 2\epsilon_{k-1})) \setminus J_{k+1}, C_p)
+
\sum_p
N(H'_{k, p}, J_{k+1}+  I(0, 3\epsilon_{k-1}))
\\
&&
+
\sum_p
N(H''_{k, p}, J_{k+1} +  I(0, 2\epsilon_{k-1})).
\end{eqnarray*}
Therefore if 
$x_{j, p}=x(E_{j,p})$, 
$y_{j, p}=x(\Phi(E_{j, p}))$, 
we have
\begin{eqnarray*}
&&
{\bf E} \left[
\sum_p
\left|
\xi_{k+1, p}^{(1)}(f) - \eta_{k+1, p}(f)
\right| ; \Omega_k 
\right]
\\
& \le &
{\bf E} \left[
\sum_p \sum_j
\left|
f(| \Lambda_{k+1} | (E_{j, p} - E_0), L_{k+1}^{-1} x_{j, p})
-
f(| \Lambda_{k+1} | (\Phi(E_{j, p}) - E_0), L_{k+1}^{-1} y_{j, p})
\right| ; \Omega_k
\right]
\\
&&+
\sum_p \| f \|_{\infty}
{\bf E} \left[
N(H_{k+1}, (J_{k+1} +  I(0, 2\epsilon_{k-1})) \setminus J_{k+1}, C_p)
\right]
\\
&&
+
\sum_p \| f \|_{\infty}
{\bf E} \left[
N(H'_{k, p}, J_{k+1} +  I(0, 3\epsilon_{k-1}))
\right]
\\
&&
+
\sum_p \| f \|_{\infty}
{\bf E}  \left[
N(H''_{k, p}, J_{k+1} +  I(0, 2\epsilon_{k-1}))
\right]
\\
&=:& I + II + III + IV.
\end{eqnarray*}
Since
$f$
is uniformly continuous, for any 
$\epsilon > 0$
we have 
$| f(x) - f(y) | < \epsilon$
whenever 
$| x - y | < \delta(\epsilon)$
with some 
$\delta(\epsilon) > 0$.
Since 
\begin{eqnarray*}
&&
\left|
\left(
| \Lambda_{k+1} | (E_{j, p} - E_0), L_{k+1}^{-1} x_{j, p}
\right)
-
\left(
| \Lambda_{k+1} | (\Phi(E_{j, p}) - E_0), L_{k+1}^{-1} y_{j, p}
\right)
\right|
\\
& \le &
| \Lambda_{k+1} |
L_{k}^d \epsilon_{k-1}
+
L_k / L_{k+1}
< \delta(\epsilon) 
\end{eqnarray*}
for large
$k$, 
we have
\begin{eqnarray*}
I & \le &
\epsilon
\,
{\bf E}\left[
\sum_p 
N(H_{k+1}, J_{k+1}, C_p)
\right]
\le
\epsilon
\, 
C_W 
\frac {2 a}{ | \Lambda_{k+1} |}
| \Lambda_{k+1} |
=
(const.) \, \epsilon
\end{eqnarray*}
by Wegner's estimate, 
which also gives a bound for 
$II$.
\begin{eqnarray*}
II & \le &
\| f \|_{\infty}
{\bf E} \left[
N(H_{k+1}, (J_{k+1} +  I(0, 2\epsilon_{k-1})) \setminus J_{k+1})
\right]
\\
& \le &
\| f \|_{\infty} C_W  4 e^{- \gamma'L_{k-1}/2} | \Lambda_{k+1} |
= o(1).
\end{eqnarray*}
$III, IV$
can be estimated similarly as in Step 2 : 
\begin{eqnarray*}
III, \, IV
& \le &
\sum_p \| f \|_{\infty}
C_W
| D_p \setminus C_p | 
\left(
\frac {2 a}{| \Lambda_{k+1} |} + 6 \epsilon_{k-1}
\right)
\\
& \le &
(const.)
\| f \|_{\infty}
\left(
\frac {L_{k+1}}{L_k}
\right)^d
C_W
L_{k-1} L_k^{d-1}
L_{k+1}^{-d} 
\\
& \le &
(const.) \frac {L_{k-1}}{L_k}.
\end{eqnarray*}
The proof of 
Proposition \ref{infinite divisibility}
is thus completed. 
\QED
\end{proof}
We next show 
some elementary bounds of 
$\xi_{k+1}, \eta_{k+1}$
to study their basic properties. 
\begin{lemma}
\mbox{}\\
(1)
For a bounded interval 
$A (\subset {\bf R} \times K)$, 
\label{pre priori estimate}
\[
{\bf E} [\xi_{k+1}(A) ] \le C_W | \pi_e (A) |, 
\quad
{\bf E} [\eta_{k+1, p}(A) ] \le C_W
\frac {| \pi_e (A) |}{| \Lambda_{k+1} |} \cdot | \Lambda_k | 
\]
for large 
$k$. \\
(2)
For 
$f \in C_c({\bf R} \times K)$
we have 
\[
{\bf E}[ \sum_p | \eta_{k+1, p}(f) | ] 
\le
(const.) C_W \| f \|_1
\]
for large 
$k$. 
\end{lemma}
\begin{proof}
(1)
Since 
$J_{k+1} := E_0 + | \Lambda_{k+1} |^{-1} \pi_e (A) \subset I$
for large
$k$, 
Wegner's estimate gives
\begin{eqnarray*}
{\bf E}[\xi_{k+1}(A)]
&\le&
{\bf E}[N(H_{k+1}, J_{k+1})] 
\le
C_W | \pi_e (A) |
\\
{\bf E}[\eta_{k+1, p}(A)]
&\le&
{\bf E}[N(H_{k, p}, J_{k+1})] 
\le
C_W \frac {| \pi_e (A) |}{| \Lambda_{k+1} |} \cdot | \Lambda_k |.
\end{eqnarray*}
(2)
Let 
$A = J \times B$
($J \subset {\bf R}, B \subset K$)
be an interval. 
Taking 
$D'_p = L_{k+1}^{-1} D_p$
we have
\[
\sum_p \eta_{k+1, p}(A)
\le
\sum_{p :  D'_p \cap B \ne \emptyset}
N(H_{k, p}, E_0 + L_{k+1}^{-d} J).
\]
Since
\[
\sharp \left\{
p : D'_p \cap B \ne \emptyset \right\}
\le (const.)
\frac {(L_{k+1}B)^d}{L_k^d}
\le
(const.) \frac { |B| \cdot | \Lambda_{k+1} |}
{| \Lambda_k |}, 
\]
we obtain, using 
Wegner's estimate again, 
\begin{eqnarray*}
{\bf E}[ \sum_p \eta_{k+1, p} (A) ] 
& \le &
(const.) 
\frac {|B| \cdot | \Lambda_{k+1} |}{ | \Lambda_k |}
\cdot 
C_W 
\frac {|J|}{| \Lambda_{k+1} |} | \Lambda_k |
= (const.) C_W |A|.
\end{eqnarray*}
A density argument gives the result. 
\QED
\end{proof}
The following lemma easily follows from 
Lemma \ref{pre priori estimate}(1).
\begin{lemma}
{\bf \mbox{}}\\
(1)
$\{ \eta_{k+1, p} \}_{p=1}^{N_k}$
is a null-array, i.e., for any bounded interval 
$A (\subset {\bf R} \times K)$, 
\[
\lim_{k \to \infty} \sup_{1 \le p \le N_k} 
{\bf P}(\eta_{k+1, p}(A) \ge 1) = 0.
\]
(2)
We have the following tightness condition
\[
\lim_{t \to \infty}
\limsup_{k \to \infty}
{\bf P}
\left(
\sum_p \eta_{k+1, p}(A) \ge t
\right)
= 0.
\]
Hence by 
\cite[Lemma 4.5]{Kallenberg}, 
$\{ \sum_p \eta_{k+1, p} \}_k$
is relatively compact. 
\label{null-array}
\end{lemma}
We sum up the results obtained in this section.
\begin{theorem}
%
Assume 
Assumption A
and 
$n(E_0) < \infty$.
Then 
$\{ \xi_k \}$
has a convergent subsequence and the limiting point 
$\xi$
is infinitely divisible whose intensity measure satisfies
\label{partial result}
\[
{\bf E}[ \xi (A) ] \le n(E_0) | A |, 
\quad
A \in {\cal B}({\bf R} \times K).
\]
\end{theorem}
\begin{proof}
The infinite divisibility 
follows from 
\cite[Theorem 6.1]{Kallenberg}, 
Proposition \ref{infinite divisibility}
and Lemma \ref{null-array}.
The claim
for the intensity measure follows from the following three considerations. 
\\
(1)
If 
$\xi_k \stackrel{d}{\to} \xi$, 
then
$\xi_k f \stackrel{d}{\to} \xi f$
for 
$f \in C_c({\bf R} \times K), f \ge 0$
\cite[Lemma 4.4]{Kallenberg}.
Hence
\[
{\bf E}[\xi (f)] \le 
\liminf_{k \to \infty} {\bf E}[\xi_{k+1}(f)].
\]
(2)
By a density argument using
Lemma \ref{pre priori estimate}(2), 
we deduce from 
(\ref{four})
(Assumption B
is not used to derive
(\ref{four}))
\[
{\bf E}[ \sum_p \eta_{k+1, p}(f) ] 
\to 
n(E_0) \| f \|_1, 
\quad
f \in C_c ({\bf R} \times K). 
\]
(3)
By
Proposition \ref{infinite divisibility}, 
\[
{\bf E}[\sum_p \eta_{k+1, p}(f)] - {\bf E}[\xi_{k+1}(f)] \to 0, 
\quad
f \in C_c ({\bf R} \times K). 
\]
\QED
\end{proof}
%
%
\section{Poisson Limit Theorem}
In this section, 
we show that 
$\{ \xi_k \}$
converges in distribution to the Poisson process, 
under Assumption A, B. 
Two conditions
(\ref{one}), (\ref{two})
in Proposition \ref{Poisson limit theorem} below are sufficient to prove that.
\begin{proposition}
%
Under 
Assumption A, B, 
we have for a bounded interval 
$A (\subset {\bf R} \times K)$, 
\label{Poisson limit theorem}
\begin{eqnarray}
(1)&&\quad
\sum_p {\bf P}(\eta_{k+1, p}(A) \ge 2) \to 0,
\label{one}
\\
(2) &&\quad
\sum_p {\bf P}(\eta_{k+1, p}(A) \ge 1) \to n(E_0) |A|.
\label{two}
\end{eqnarray}
\end{proposition}
Proposition \ref{Poisson limit theorem}
together with 
\cite[Corollary 7.5]{Kallenberg}, 
Proposition \ref{infinite divisibility}
and
Lemma \ref{null-array}
proves Theorem \ref{main theorem}.
For its proof, a preparation is necessary. 
\begin{lemma}
%
Assume
Assumption A.
For an interval
$J (\subset {\bf R})$, 
we have
\label{Step 5}
\[
\sum_{p=1}^{N_k} {\bf E}[\eta_{k+1, p}(J \times K)]
\to n(E_0) |J|.
\]
\end{lemma}
\begin{proof}
Since 
$
{\bf E}[ | \xi_{k+1}(J \times K) - \sum_p \eta_{k+1, p}(J \times K) | ] \to 0
$
by Proposition \ref{infinite divisibility}
and 
Lemma \ref{pre priori estimate}(1), 
it suffices to show 
\[
{\bf E}[ \xi_{k+1}(J \times K) ] \to n(E_0) | J |.
\]
As is done in \cite{Minami}, 
it is further sufficient to show the above equation for the following function instead of 
$1_J$
\[
f_{\zeta}(x) = \frac {\tau}{(x -\sigma)^2 + \tau^2},
\quad
\zeta = \sigma + i \tau \in {\bf C}_+, 
\]
because the set 
\[
{\cal A} := \left\{
\sum_{j=1}^n a_j f_{\zeta_j} (x) : 
\;
a_j \ge 0, \; \zeta_j \in {\bf C}_+
\right\}
\]
of the finite linear combinations of 
$f_{\zeta}$
with positive coefficients is dense in 
$L^1_+({\bf R})$
\cite[Lemma 1]{Minami}, 
and 
Lemma \ref{a priori estimate}
then enables us to carry out the density argument. 
Hence it suffices to show
\[
{\bf E}[\xi_{k+1}(f_{\zeta})] \to \pi n(E_0),
\quad
\zeta \in {\bf C}_+.
\]
For any 
$x \in \Lambda_{k+1}$,
we have
\begin{eqnarray*}
{\bf E}[ \xi_{k+1}(f_{\zeta}) ] 
&=&
\frac {1}{| \Lambda_{k+1} |}
{\bf E} [ \mbox{Tr } \Im G_{k+1}(E_0 + \frac {\zeta}{| \Lambda_{k+1} |}) ]
=
{\bf E}[ \Im G_{k+1}(E_0 + \frac {\zeta}{| \Lambda_{k+1} |}; x,x) ],
\end{eqnarray*}
since $H_{k+1}$ has periodic b.c. 
Let 
$G(z)=(H - z)^{-1}, G_{k+1}(z) = (H_{k+1} - z)^{-1}$
be Green's function of 
$H, H_{k+1}$
respectively. 
Let 
$x$
be the center of 
$\Lambda_{k+1}$
and let 
$z_{k+1} = E_0 + \frac {\zeta}{| \Lambda_{k+1} |}$.
Then by the resolvent equation, 
\begin{eqnarray*}
&&
| G_{k+1}(z_{k+1};x,x) - G(z_{k+1};x,x) |
\\
&&\quad
\le
\sum_{\langle y, y' \rangle \in \tilde{\partial} \Lambda_{k+1}
\cup
\bar{\partial} \Lambda_{k+1}}
| G_{k+1}(z_{k+1};x,y) 
G(z_{k+1} ; y', x) |
\end{eqnarray*}
where
$\langle y, y' \rangle \in \bar{\partial} \Lambda_{k+1}$
means that 
$y' \in \partial \Lambda_{k+1}$
is connected to 
$y \in \partial \Lambda_{k+1}$
if we regard 
$\Lambda_{k+1}$ 
as a torus.
By the multiscale analysis,  the event
\begin{equation}
{\cal G}_{k+1}(E) 
:=
\left\{
\omega \in \Omega : 
\mbox{ $\Lambda_{k+1}$
is 
$(\gamma_0, E)$-regular }
\right\}
\label{Gk}
\end{equation}
satisfies
\begin{equation}
{\bf P}\left(
{\cal G}_{k+1}(E)
\right)
\ge 1 - L_{k+1}^{-p}
\label{peribc}
\end{equation}
for any 
$E \in I$
and 
$0 < \gamma_0 < \gamma$. 
Although 
$H_{k+1}$
has periodic b.c., the procedure of the multiscale analysis requires no essential modifications to prove (\ref{peribc}). 
Take 
$k$
large enough and let
$E_{k+1} = \Re z_{k+1}$. 
We decompose
\begin{eqnarray*}
&&
\left|
{\bf E}[ G_{k+1}(z_{k+1} ; x,x)
-
{\bf E}[ G(z_{k+1} ; x,x) ]
\right|
\\
& \le &
\sum_{\langle y, y' \rangle \in \tilde{\partial} \Lambda_{k+1}
\cup
\bar{\partial} \Lambda_{k+1}}
{\bf E} [
| G_{k+1}(z_{k+1} ; x,y) |
\cdot
| G(z_{k+1} ; y', x) |
 ; 
{\cal G}_{k+1}(E_{k+1}) ]
\\
&&\qquad
+
\sum_{\langle y, y' \rangle \in \tilde{\partial} \Lambda_{k+1}
\cup
\bar{\partial} \Lambda_{k+1}}
{\bf E} [
| G_{k+1}(z_{k+1} ; x,y) | 
\cdot
| G(z_{k+1}; y',x ) |; 
{\cal G}_{k+1}(E_{k+1})^c ]
\\
&=:& I + II.
\end{eqnarray*}
By
(\ref{peribc}), we have
\begin{eqnarray*}
I & \le &
c_d L_{k+1}^{d-1} e^{- \gamma_0 \frac {L_{k+1}}{2}}
L_{k+1}^d
= o(1),
\quad
II  \le 
c_d L_{k+1}^{d-1} L_{k+1}^{2d} L_{k+1}^{-p},
\end{eqnarray*}
so that 
$p > 3d - 1$
is required to have
$II = o(1)$,
which is guaranteed by 
Assumption A(1). 
Therefore
\begin{eqnarray*}
{\bf E}[ \xi_{k+1}(f_{\zeta}) ] 
&=&
{\bf E}[ \Im G(z_{k+1}; x, x)]
+ o(1)
=
\pi n(E_0) + o(1).
\end{eqnarray*}
as
$k \to \infty$. 
\QED
\end{proof}
{\it Proof of Proposition \ref{Poisson limit theorem} }
Let 
$A (\subset {\bf R} \times K)$
be a bounded interval. 
As is discussed in 
\cite{Minami}, 
it suffices to show the following equations to prove
Proposition \ref{Poisson limit theorem}.
\begin{eqnarray}
(1) &&\quad
\sum_{j \ge 2} \sum_p 
{\bf P}( \eta_{k+1, p}(A) \ge j) \to 0,
\label{three}
\\
(2) &&\quad
\sum_p {\bf E}[ \eta_{k+1, p} (A) ] \to n(E_0) |A|.
\label{four}
\end{eqnarray}
In fact, 
(\ref{three})
trivially implies
(\ref{one}), 
and
(\ref{two})
follows from 
\[
\sum_p {\bf P}
(\eta_{k+1, p}(A) \ge 1)
=
\sum_p {\bf E}[\eta_{k+1, p}(A)]
-
\sum_p \sum_{j \ge 2}
{\bf P}(\eta_{k+1, p}(A) \ge j)
\to
n(E_0) |A|.
\]
(\ref{three})
in turn follows from 
Assumption B(Minami's estimate) : 
\begin{eqnarray*}
\sum_p 
\sum_{j \ge 2}
{\bf P}(\eta_{k+1, p}(A) \ge j)
& \le &
\sum_p \sum_{j \ge 2}
j(j-1)
{\bf P}(\eta_{k+1, p}(\pi_e(A) \times K) = j)
\\
& \le &
C_M
N_k \frac {| \Lambda_k |^2}{| \Lambda_{k+1} |^2}
\to 0
\end{eqnarray*}
which is the only (and fundamental) step to use
Minami's estimate.

To prove
(\ref{four}), 
let 
$J = \pi_e (A), B = \pi_s (A)$
and let
$D'_p = L_{k+1}^{-1} D_p$.
We then have
\begin{eqnarray}
\sum_p {\bf E}[ \eta_{k+1, p}(A) ]
&=&
\sum_{D'_p \subset B}
{\bf E}[ \eta_{k+1, p}(A) ]
+
\sum_{B \cap D'_p \ne \emptyset, 
B \cap D_p^{'c} \ne \emptyset}
{\bf E}[ \eta_{k+1, p}(A) ]
\nonumber
\\
&=:& I + II.
\label{first}
\end{eqnarray}
By
Lemma \ref{pre priori estimate}(1)
and by the inequality
$
\sharp \{ p : 
D_p' \cap B \ne \emptyset, 
D_p' \cap B^c \ne \emptyset \}
\le
(const.)
\left(
\frac {L_{k+1}}{L_k}
\right)^{d-1}, 
$
we have
\begin{eqnarray}
II
& \le &
(const.)
\left(
\frac {L_{k+1}}{L_k}
\right)^{d-1}
\cdot
\frac 
{| \Lambda_k |}{| \Lambda_{k+1} |}
=
(const.)
\frac {L_k}{L_{k+1}}.
\label{second}
\end{eqnarray}
To compute 
$I$, 
we note
$
I 
=
\sharp \{ p : D'_p \subset B \}
{\bf E}[ \eta_{k+1, p}(J \times K) ].
$
Substituting 
$
N_k {\bf E}[\eta_{k+1, p}(J \times K)]
=
n(E_0) | J | + o(1),
$
which follows from Lemma \ref{Step 5}, we have
\begin{equation}
I
=
\frac {\sharp \{ p : D'_p \subset B \}}
{N_k}
\left(
n(E_0) | J | + o(1) 
\right)
=
n(E_0) |B| \cdot |J| + o(1)
\label{third}
\end{equation}
as
$k \to \infty$. 
By
(\ref{first}), (\ref{second})
and 
(\ref{third}), 
we obtain
(\ref{four}).
\QED
%
\section{Appendix 1}
%
\subsection{Basic properties of localization centers}
We review some basic properties of localization centers\cite{BG, KM}. 
\begin{lemma}\cite{BG}
%
Let 
$H \phi = E \phi, \phi \in l^2 ({\bf Z}^d)$
($H = H_{k+1}$
or
$H=H_{k, p}$).
Then we can find
$L_0 (d, \gamma)$
such that for 
$L \ge L_0$, 
$\Lambda_L (x(\phi))$
(with Dirichlet b.c.) is
$(\gamma, E)$-singular.
\label{loc ctr lies in singular box}
\end{lemma}
\begin{lemma}
%
For any 
$0 <  \gamma_m < \gamma$
we can find 
$k_1=k_1(\alpha, d, \gamma, \gamma_m)$
with the following properties.
Suppose
$\omega \in \Omega_k$
and
$\phi \in {\cal E}f(H_{k+1}, I, C_p)$
for some 
$p = 1, 2, \cdots, N_k$.
Then if
$k \ge k_1$
we have
\label{phi decays at boundary}
\begin{eqnarray*}
&&
\| ( 1 - \chi_{D_p}) \phi \|_{l^2(\Lambda_{k+1})}
\le
e^{- \gamma_m \frac {L_{k-1}}{2}}.
\end{eqnarray*}
$D_p, C_p$
are defined in Section 2. 
\end{lemma}
\begin{proof}
Take 
$k_1$
large enough with 
$L_{k_1} \ge L_0(d, \gamma)$.
Since
$\omega \in \Omega_k$
and since
$\Lambda_{k-1}(x(\phi))$
is
$(\gamma, E)$-singular
by Lemma \ref{loc ctr lies in singular box}, 
$\Lambda_{k-1}(x)$
(with Dirichlet b.c.) is
$(\gamma, E)$-regular
for
$x \notin D_p$.
Here, as in (\ref{eventk}),  we regard
$\Lambda_{k+1}$
as a torus and 
$\Lambda_{k-1}(x) \subset \Lambda_{k+1}$.
Therefore, using 
$| \phi (y) | \le 1$ 
we have
\begin{eqnarray*}
| \phi (x) | 
& \le &
\sum_{\langle y, y' \rangle \in \tilde{\partial} \Lambda_{k-1}(x)}
| G_{\Lambda_{k-1} (x)}(E; x, y)|  | \phi (y') |
\le 
c_d L_{k-1}^{d-1} e^{- \gamma \frac {L_{k-1}}{2}}.
\end{eqnarray*}
Taking 
$k_1 (\alpha, d, \gamma, \gamma_m)$
large enough with 
$L_{k_1+1}^d c_d^2 L_{k_1-1}^{2(d-1)} 
e^{ - \gamma L_{k-1}}
\le
e^{- \gamma_m L_{k-1}}$
gives the result.
\QED
\end{proof}
The following lemma 
says two localization centers 
$x(\phi), \langle x \rangle_{\phi}$
are close in the scale of 
$L_k$. 
\begin{lemma}
\label{two notions}
Let
$\omega \in \Omega_k$, 
$\phi \in {\cal E}f(H_{k+1}, I)$.
Then for 
$k \ge k_1(\alpha, d, \gamma, \gamma_m)$, 
we have
\[
| \langle x \rangle_{\phi} - x(\phi) | 
\le
L_{k-1} + (const.) e^{- \gamma'' L_{k-1}}, 
\quad
0 < \gamma'' < \gamma_m. 
\]
\end{lemma}
\begin{proof}
Set
$
A_k := \left\{
x \in \Lambda_k : 
d (x (\phi),x) \le  L_{k-1} 
\right\}.
$
Since by Lemma \ref{phi decays at boundary}, 
$\sum_{x \in A_k^c} | x | | \varphi (x) |^2
\le
(const.) e^{- \gamma'' L_{k-1}}$, 
$0 < \gamma'' < \gamma_m$, 
we have
\begin{eqnarray*}
\left|
\langle x \rangle_{\phi} - x(\phi)
\right|
& \le &
\sum_{x \in A_k} | x - x_{\phi} | | \phi (x) |^2
+
\sum_{x \in A_k^c} | x - x_{\phi} | | \phi (x) |^2
\\
& \le &
L_{k-1} + (const.) e^{- \gamma'' L_{k-1}}.
\end{eqnarray*}
\QED
\end{proof}
%

\subsection{Comparison of eigenvalues of 
big and small boxes}
In Section 2, 
we need to show that  eigenvalues of 
$H_{k+1}$
localized in 
$C_p$
produce those of 
$H_{k, p}$. 
The following lemma is an elementary extension of
\cite[Lemma 1]{KM}.
\begin{lemma}
%
For any 
$0 < \gamma' < \gamma$, 
we can find
$k_2(\alpha, d, \gamma, \gamma')$
with the following properties.
Let
$J(\subset I)$
be an interval, 
$\omega \in \Omega_k$
and
$k \ge k_2$. 
Then
\label{many eigenvalues}
\begin{eqnarray*}
(1) &&\;
N(H_{k+1}, J, C_p) \le N(H_{k, p}, J+I(0, \epsilon_{k-1}))
\\
(2) &&\;
N(H_{k+1}, J, D_p \setminus C_p) 
\le N(H'_{k, p}, J+I(0, \epsilon_{k-1}))
\\
(3) &&\;
N(H_{k, p}, J, D_p \setminus C_p) 
\le N(H''_{k, p}, J+I(0, \epsilon_{k-1}))
\end{eqnarray*}
where
$\epsilon_{k-1} := e^{- \gamma' L_{k-1}/2}$.
$D_p, C_p, H_{k, p}, H'_{k, p}$
 and 
 $H''_{k, p}$
are defined in Section 2. 
\end{lemma}
\begin{proof}
It is sufficient to show
(1).
Let 
$\{ \phi_j \}_{j=1}^{M_p} :={\cal E}f(H_{k+1}, J, C_p)$, 
$M_p := N(H_{k+1}, J, C_p)$
and set
$\psi_j = \chi_{D_p} \phi_j$.
Letting
$\gamma_m = \frac {\gamma + \gamma'}{2}$, 
we have by 
Lemma \ref{phi decays at boundary}, 
\begin{eqnarray}
&&\| \psi_j \|^2_{l^2(D_p)}
\ge
1 - e^{- \gamma_m L_{k-1}},
\label{j}
\\
&&
| \langle \psi_i, \psi_j \rangle_{l^2(D_p)} |
\le
e^{- \gamma_m L_{k-1}}, 
\quad
i, j = 1, 2, \cdots, M_p, 
\;
i \ne j
\label{ij}
\end{eqnarray}
for 
$k \ge k_1(\alpha, d, \gamma, \gamma_m)$. 
By 
(\ref{j})
and
(\ref{ij}), 
it is straightforward to prove the following Claim.\\

\noindent
{\bf Claim}\\
(1)
We can find
$k'(\alpha, d, \gamma_m)$
such that if
$k \ge k'$, 
$\psi_1, \cdots, \psi_{M_p}$
are linearly independent.\\
(2)
$\| (H_{k, p} - E_j) \psi_j \|_{l^2(D_p)}
\le
\sqrt{2}
e^{- \gamma_m L_{k-1}/2}$, 
$j = 1, 2, \cdots, M_p$. 
\\

Let
$J' := J + I(0, \epsilon_{k-1})$, 
let 
$P$
be the spectral projection of 
$H_{k, p}$
corresponding to 
$J'$
and let
$Q = I - P$. 
Since
$
\| (H_{k, p} - E_j) Q \psi_j \|_{l^2(D_p)}^2
\ge
\epsilon_{k-1}^2 
\| Q \psi_j \|_{l^2(D_p)}^2
$
by the spectral theorem, we have
\[
\| Q \psi_j \|_{l^2(D_p)}
\le
\sqrt{2} e^{- (\gamma_m -\gamma')L_{k-1}/2}, 
\quad
j = 1, 2, \cdots, M_p
\]
by Claim (2). 
Let 
$V := \mbox{ Span } \{ \psi_1, \cdots, \psi_{M_p} \}$
and take
$\psi \in V, \| \psi \|_{l^2(D_p)} = 1$. 
Writing 
$\psi = \sum_j a_j \psi_j$,
we have
\begin{equation}
1 = \| \psi \|_{l^2(D_p)}^2
=
\sum_j | a_j |^2 \| \psi_j \|_{l^2(D_p)}^2
+
\sum_{i \ne j}
a_i \overline{a_j}
\langle \psi_i , \psi_j \rangle_{l^2(D_p)}.
\label{norm}
\end{equation}
By inequalities
(\ref{j}), (\ref{ij}) and 
\begin{eqnarray*}
\left|
\mbox{ 2nd term of (\ref{norm}) }
\right|
\le
e^{- \gamma_m L_{k-1}} \sum_{i \ne j} | a_i | | a_j |
\le
e^{- \gamma_m L_{k-1}} (M_p - 1) 
\sum_i | a_i |^2, 
\end{eqnarray*}
we have
$\sum_j | a_j |^2
\le
( 1 - M_p e^{- \gamma_m L_{k-1}} )^{-1}$
and hence
\begin{eqnarray*}
\| Q \psi \|_{l^2(D_p)}^2
& \le &
\sum_j | a_j |^2 
\cdot
\sum_j \| Q \psi_j \|_{l^2(D_p)}^2
\le
\frac {2 | \Lambda_{k+1} | e^{- (\gamma_m - \gamma')L_{k-1}}}
{1 - | \Lambda_{k+1} | e^{- \gamma_m L_{k-1}}}.
\end{eqnarray*}
Taking 
$k \ge k_2 (\alpha, d, \gamma, \gamma')$
such that
$\frac {2 | \Lambda_{k+1} | e^{- (\gamma_m - \gamma')L_{k-1}}}
{1 - | \Lambda_{k+1} | e^{- \gamma_m L_{k-1}}}
< \frac 12$, 
we have
$\| Q \psi \|_{l^2(D_p)}^2 < \frac 12 \| \psi \|_{l^2(D_p)}^2$
and hence
\[
\| P \psi \|_{l^2(D_p)}^2
> 
\frac 12
\| \psi \|_{l^2(D_p)}^2
\]
which implies
$P$
is injective on 
$V$.
Therefore
$\dim \mbox{ Ran } P \ge \dim PV = M_p$. 
\QED
\end{proof}
We next do the converse : 
we show that an eigenvalues of 
$H_{k, p}$
localized in 
$C_p$
produce those of 
$H_{k+1}$. 
Since the proofs are similar to those of Lemma \ref{phi decays at boundary} and \ref{many eigenvalues}, we state the result only. 
\begin{lemma}
\label{many eigenvalues3}
For any 
$0 < \gamma' < \gamma$, 
we can find
$k_3 = k_3 (\alpha, d, \gamma, \gamma')$
with the following property.
Suppose
$\omega \in\Omega_k$, 
$J (\subset I)$
is an interval and 
$k \ge k_3$, 
then 
\[
\sum_p 
N(H_{k, p}, J, C_p) \le N(H_{k+1}, J+I(0, \epsilon_{k-1}).
\]
\end{lemma}
%
%
\subsection{A priori estimate}
We show a priori estimate for 
${\bf E}[| \xi_{k+1}(g) |]$
for 
$g(x) = O(|x|^{-2})$ 
as
$|x| \to \infty$. 
\begin{lemma}
\label{a priori estimate}
Suppose 
$g$
is bounded and measurable on 
${\bf R}$, 
satisfying
\[
| g(x) | \le \frac {C_R}{x^2}, 
\quad
|x| \ge R
\]
for some 
$R>0$
and 
$C_R > 0$. 
Let
$r := d (E_0, I^c) > 0$.
If 
$r | \Lambda_{k+1} | \ge R$, 
we have
\[
{\bf E} [ | \xi_{k+1}(g) | ]
\le
C_W \int_{\{ |\lambda | < r | \Lambda_{k+1} | \}} | g(\lambda) | d \lambda
+
\frac {C_R}{r^2 | \Lambda_{k+1} |}.
\]
\end{lemma}
\begin{proof}
We decompose
\begin{eqnarray*}
\xi_{k+1}(g)
&=&
\sum_{E_j \in I}g(| \Lambda_{k+1} | (E_j(\Lambda_{k+1}) - E_0)) 
+
\sum_{E_j \in I^c}g(| \Lambda_{k+1} | (E_j(\Lambda_{k+1}) - E_0)) 
\\
&=& I + II.
\end{eqnarray*}
$II$
is estimated by using the assumption on 
$g$. 
\begin{equation}
| II |
\le
| \Lambda_{k+1} | \cdot
\frac {C_R}{r^2 | \Lambda_{k+1} |^2}.
\label{estimate for II}
\end{equation}
To estimate 
$I$, 
we note
$
I
=
\xi_{k+1}
( g 1_{\{ | \lambda | < r | \Lambda_{k+1} | \}})
$.
If 
$g = 1_J$
for some interval
$J \subset \{ x \in {\bf R} : |x| < r| \Lambda_{k+1} | \}$, 
we have by Wegner's estimate, 
\begin{eqnarray}
{\bf E} [ \xi_{k+1} (g 1_{\{ |x| < r | \Lambda_{k+1} | \}}) ]
&=&
{\bf E} [N(H_{k+1}, E_0 + \frac {J}{| \Lambda_{k+1}|})]
\nonumber
\\
&\le&
C_W | J |
=
C_W 
\int_{ \{ |\lambda | < r | \Lambda_{k+1} | \}} 
| g (\lambda) | d \lambda.
\label{estimate for g}
\end{eqnarray}
A density argument 
proves 
(\ref{estimate for g})
for 
$g$
bounded and measurable. 
Together with
(\ref{estimate for II}), we arrive at the conclusion.
\QED
\end{proof}
%
\section{Appendix 2}
In this section, 
we consider the random measure 
$\xi$ 
studied in 
\cite{Killip-Nakano}, 
and examine its natural scaling limit under Assumption A. 
$\xi$
is defined by 
\[
\xi (J \times B) := \mbox{Tr }
(1_B (x) P_J (H))
\]
for an interval 
$J \times B$ $(J \subset {\bf R}, B \subset {\bf R}^d)$, 
and its scaling 
$\xi_L$
is given by
\[
\xi_L (J \times B) :=
\mbox{Tr }\left(
1_{LB} (x) P_{E_0 + L^{-d} J} (H)
\right), 
\quad
L > 0
\]
which is done in the same spirit of 
$\xi_k$. 
$P_J(H)$
is the spectral projection of 
$H$
corresponding to 
$J$. 
We then have
\begin{theorem}
%
\label{main theorem2}
Suppose 
Assumption A
(with 
$p > 8 d - 2$)
and 
$n(E_0) < \infty$. 
Then we can find a convergent subsequence
$\{ L_k \}_{k=1}^{\infty}$
such that 
$\xi_{L_k}$
converges in distribution  to a infinitely divisible point process
$\xi$ 
on 
${\bf R}^{d+1}$
with its intensity measure satisfying
\[
{\bf E} \xi (dE \times dx)
\le
n(E_0) d E \times dx.
\]
\end{theorem}
For its proof, we take
$l_L = O(L^{\beta})$
for some 
$0 < \beta < 1$
and consider 
\begin{eqnarray*}
B_p (L) &:=& \left\{
x \in {\bf Z}^d : p_j l_L \le x_j < (p_j+1) l_L, 
\;
j = 1, \cdots, d \right\},
\quad
p \in {\bf Z}^d
\\
H_{L,p} & := & H |_{B_p(L)},
\quad
H_L := \oplus_p H_{L, p}
\end{eqnarray*}
as is done in \cite{Killip-Nakano}, 
with the periodic boundary condition. 
Let 
$\tilde{\eta}_{L, p}$
be the random measure defined by
\[
\tilde{\eta}_{L, p}(J \times B) :=
\mbox{Tr }\left(
1_{LB} (x) P_{E_0 + L^{-d} J} (H_{L, p})
\right).
\]
We then have
\begin{proposition}
%
Suppose 
Assumption A with
$p > 8d - 2$. 
Then for
$f \in C_c({\bf R}^{d+1})$, 
\label{infinite divisibility2}
\begin{equation}
{\bf E} [
|
\xi_L (f) - \sum_p \tilde{\eta}_{L, p}(f)
|
]
\to 0.
\label{KN}
\end{equation}
\end{proposition}
{\it Sketch of proof of Proposition \ref{infinite divisibility2}}\\
{\it Step 1} : 
We first show
(\ref{KN})
for 
$f(E,x) = 1_B(x) f_{\zeta}(E)$
for a box 
$B \subset {\bf Z}^d$
and 
$\zeta \in {\bf C}_+$.
$f_{\zeta}$
is defined in Section 3. 
In order to do that, 
we use the resolvent equation, decompose the expectation into good and bad events, and use the following estimate given by the multiscale analysis : 
\begin{equation}
{\bf P} \left(
\sup_{\epsilon > 0}
| G_{\Lambda} (E+ i \epsilon ; x , y)| 
\le
e^{- \frac {\gamma}{8}  |x-y|}
\right)
\ge 
1 - C | \Lambda | |x - y|^{- p/2}
\label{MSA for any boxes}
\end{equation}
for any 
$E \in I$, 
any box
$\Lambda$
($H_{\Lambda}$
has periodic b.c.) 
and any 
$x ,y \in \Lambda$
with 
$|x - y| \ge C$
for some 
$C$.\\
{\it Step 2} : 
We prove a simple estimate
\begin{equation}
{\bf E} \left[
\left|
\int 1_B (x) g(E) d \xi_L 
\right|
\right]
\le
C_W (1 + o(1)) |B| \| g \|_1
+
\frac {(const.)}{L^d}
\label{simple estimate}
\end{equation}
for 
$g$
bounded and measurable with
$|g(x)| \le \frac {C_R}{x^2}$, 
$|x| \ge R$
for some 
$R > 0$
and
$C_R > 0$. 
The estimate
(\ref{simple estimate})
can be proved similarly as
Lemma \ref{Step 5}
and 
Lemma \ref{a priori estimate}.
By a density argument using
(\ref{simple estimate}), 
we can show
(\ref{KN})
for 
$f(E,x) = 1_B (x) g(E)$
for a box
$B \subset {\bf Z}^d$
and
$g \in C_c({\bf R})$. 
Then
we can further extend 
(\ref{KN})
to arbitrary
$f \in C_c({\bf R}^{d+1})$
by using some a priori estimates stated below : 
for any 
$C > 0$
we can find
$L_0(C)$
with
\begin{eqnarray}
&&(1) \qquad
{\bf E} \left[ \left|
\int f(E,x) d \xi_L
\right| \right]
\le
2
n(E_0) \| f\|_1
\label{stated below}
\\
&&(2) \qquad
{\bf E} \left[ 
\sum_p
\left|
\int f(E,x) d \tilde{\eta}_{L, p}
\right| \right]
\le
C_W \| f\|_1.
\nonumber
\end{eqnarray}
for
$\mbox{ supp } f \subset \{ | (E, x) | \le C \}$
and
$L \ge L_0 (C)$.
An alternative way 
to prove
Proposition \ref{infinite divisibility2} is 
to use the almost analytic extensions which also applies to the continuum analog of this statement. 
\QED

The facts that the sequence
$\{ \xi_L \}_L$
is a null-array and relatively compact
follow from 
(\ref{stated below}), 
and then
Proposition \ref{infinite divisibility2} proves the infinite divisibility of the limiting random measure
$\xi$. 
The infinite divisibility of 
$\xi$
as a point process and the estimate for its intensity measure
${\bf E} \xi(dE \times dx)$ 
follow similarly as in 
\cite{Killip-Nakano},
completing the proof of Theorem \ref{main theorem2}.

\begin{remark}
Let
$B(\subset {\bf Z}^d)$
be a finite box and let 
$H_{LB} := H |_{LB}$
be a restriction of 
$H$
on 
$LB$
with some boundary condition. 
Define a random measure 
$\xi_{L, B}$
on 
${\bf R} \times B$ 
by 
\[
\xi_{L, B}(J \times C) =
\mbox{Tr }\left(
1_{LC}(x) P_{E_0 + L^{-d} J}(H_{LB})
\right), 
\quad
J \subset {\bf R}, 
\;
C \subset B. 
\]
Then for 
$f \in C_c({\bf R} \times B)$,
the proof of 
Proposition \ref{infinite divisibility2}
tells us that 
$\xi_L(f) - \xi_{L,B}(f) \stackrel{v}{\to} 0$
a.s.
Therefore 
the eigenvalues and 
the eigenfunctions on
$H_{LB}$
and those of
$H$
localized in 
$LB$
has the same behavior in this sense. 
\end{remark}

\section{Appendix 3}
In this section
we assume both
Assumption A and B, and present another presentation of Step 3 in the proof of Proposition \ref{infinite divisibility} : we show the following equation for 
$f \in C_c({\bf R} \times K)$. 
\begin{equation}
\sum_p
{\bf E} \left[
| \xi_{k+1, p}^{(1)}(f) - \eta_{k+1, p}(f) | ; \Omega_k 
\right] = o(1).
\label{sufficient}
\end{equation}
For simplicity, let 
\begin{eqnarray*}
J'_{k+1} & := & J_{k+1} + I(0, \epsilon_{k-1}).
\end{eqnarray*}
and decompose the LHS of 
(\ref{sufficient})
as
\begin{eqnarray*}
\mbox{ LHS of }(\ref{sufficient}) 
&=&
\sum_p {\bf E} \left[
| \xi_{k+1, p}^{(1)}(f) - \eta_{k+1, p}(f) | 
;
\Omega_k \cap \{ N(H_{k, p}, J'_{k+1}) = 1 \}
\right]
\\
&& +
\sum_p {\bf E} \left[
| \xi_{k+1, p}^{(1)}(f) - \eta_{k+1, p}(f) | 
;
\Omega_k \cap \{ N(H_{k, p}, J'_{k+1}) \ge 2  \}
\right]
\\
&=:& A + B. 
\end{eqnarray*}

\noindent
{\bf Claim 1 : }
$B = o(1)$. \\

{\it Proof of Claim 1 }
We write
$B = \sum_p B_p$. 
By 
Lemma \ref{many eigenvalues}(1)
and by Minami's estimate, we have
\begin{eqnarray*}
B_p
& \le &
2\| f \|_{\infty}
{\bf E} \left[
N(H_{k, p}, J'_{k+1}) ; 
\Omega_k \cap \{ N (H_{k, p},J'_{k+1}) \ge 2 \}
\right]
\\
& \le &
2 \| f \|_{\infty}
\sum_{j \ge 2} j (j-1) 
{\bf P} \left( N (H_{k, p},J'_{k+1}) = j \right)
\\
& \le &
2 \| f \|_{\infty} C_M
\left(
\frac {2a}{| \Lambda_{k+1}|} + 2 \epsilon_{k-1}
\right)^2
\cdot
| \Lambda_k |^2
\end{eqnarray*}
which shows 
$
B
\le
(const.) 
\frac {| \Lambda_k |}{| \Lambda_{k+1} |}
$
and thus proves 
Claim 1.
\QED
\\

To estimate
$A$, 
we further decompose
$A = A_1 + A_2$
with
\begin{eqnarray*}
A_1 
&=&
\sum_p {\bf E} \left[
| \eta_{k+1, p}(f) - \xi_{k+1, p}^{(1)}(f) | ; 
\Omega_k \cap 
\{ N(H_{k, p}, J'_{k+1}) = 1, 
\;
N(H_{k+1}, J_{k+1}, C_p) = 1 \}
\right]
\\
A_2
&=&
\sum_p {\bf E} \left[
| \eta_{k+1, p}(f) | ; 
\Omega_k \cap 
\{ N(H_{k, p}, J'_{k+1}) = 1, 
\;
N(H_{k+1}, J_{k+1}, C_p) = 0 \}
\right].
\end{eqnarray*}
Here we note that 
$| \xi_{k+1, p}^{(1)}(f) - \eta_{k+1, p}(f) | = 0$
if 
$N(H_{k, p}, J'_{k+1}) = 0$
by
Lemma \ref{many eigenvalues}(1).
\\

\noindent
{\bf Claim 2 : }
$A_2 = o(1)$.
\\

{\it Proof of Claim 2 }
Lemma \ref{bound for eta}
and the argument 
in the proof of Claim 1 gives
\begin{eqnarray}
{\bf E}[ N(H_{k+1}, J_{k+1}) ] 
&=&
\sum_p {\bf E} \left[
N(H_{k+1}, J_{k+1}, C_p) ; 
\Omega_k \cap 
\{ N(H_{k, p}, J'_{k+1}) = 1 \}
\right]
+ o(1)
\qquad\quad
\label{sharp}
\\
{\bf E}[ N(H_{k+1}, J_{k+1}) ] 
&=&
\sum_p {\bf E}
[ N(H_{k, p}, J'_{k+1}) ] + o(1)
\nonumber
\\
&=&
\sum_p {\bf E} \left[
N(H_{k, p}, J'_{k+1}) ; 
\Omega_k \cap \{ N(H_{k, p}, J'_{k+1}) = 1 \} 
\right]
+ o(1).
\qquad\quad
\label{natural}
\end{eqnarray}
By
Lemma \ref{many eigenvalues}(1), (\ref{sharp}) and (\ref{natural}), we have
\begin{eqnarray}
0 
\le
\sum_p {\bf E} \left[
N(H_{k,p}, J'_{k+1}) - N(H_{k+1}, J_{k+1}, C_p) ; 
\Omega_k \cap \{ N(H_{k, p}, J'_{k+1}) = 1 \}
\right]
= o(1).
\qquad\quad
\label{asterisque}
\end{eqnarray}
Since we have
\begin{eqnarray*}
| \eta_{k+1, p}(f) |
& \le &
\| f \|_{\infty}
\left(
N(H_{k, p}, J'_{k+1}) - N(H_{k+1}, J_{k+1}, C_p) 
\right)
\end{eqnarray*}
on the event in which
$A_2$
is computed, 
(\ref{asterisque})
implies
$A_2 = o(1)$
and thus proves Claim 2. 
\QED\\

On the event in which
$A_1$
is computed, it is easy to construct bijective correspondence between 
${\cal E}(H_{k+1}, J_{k+1}, C_p)$
and
${\cal E}(H_{k, p}, J'_{k+1})$
which proves
Proposition \ref{infinite divisibility}.
It remains to show the following lemma.
\begin{lemma}
\label{bound for eta}
If 
$p > 12 d$
in Assumption A(1), we have
\[
{\bf E}[ N(H_{k+1}, J_{k+1}) ]
=
\sum_p {\bf E} 
[N (H_{k, p}, J'_{k+1})] + o(1).
\]
\end{lemma}
\begin{proof}
By Wegner's estimate, it suffices to show
${\bf E}[ N(H_{k+1}, J_{k+1}) ]
=
\sum_p {\bf E} 
[N (H_{k, p}, J_{k+1})] + o(1)$.
By
Lemma \ref{pre priori estimate}, 
it is further reduced to 
${\bf E}\left[
\left| 
\xi_{k+1}(f) - \sum_p \eta_{k+1, p}(f) 
\right|
\right]
= o(1)$
for any 
$f \in C_c ({\bf R})$.
By the density of 
${\cal A}$
in
$L_+^1({\bf R})$, 
it is sufficient to take
$f = f_{\zeta}$, $\zeta \in {\bf C}_+$
in which case the proof can be done by using the resolvent equation and the exponential decay of Green's functions (\ref{MSA for any boxes}). 
\QED
\end{proof}
%
\section{Appendix 4 : Proof of Theorem 1.1
}
To prove 
Theorem \ref{uniform distribution}, it suffices to show
\begin{equation}
\bar{\xi}_k (J \times B) \to \nu (J) |B|, \quad a.s.
\label{suffice}
\end{equation}
for intervals
$J \subset I, B \subset K$
with rational endpoints. 
Let
$B'_k := (L_k B) \cap {\bf Z}^d$.
Then 
$| B'_k | = |B| L_k^d (1 + o(1))$
for large 
$k$
and
\begin{equation}
\bar{\xi}_k (J \times B)
=
\frac {1}{| \Lambda_k |}
N(H_k, J, B'_k).
\label{equality}
\end{equation}
We also consider a box
$D_k$
by eliminating a strip of width 
$2 L_{k-1}$ 
from the boundary of 
$B'_k$
and further consider boxes
$B''_k$
(resp. $B'''_k$)
obtained by adding a strip of width
$L_{k-1}$
in both sides of the strip
$B_k' \setminus D_k$
in 
$\Lambda_k$
(resp. in $B_k'$) : 
\begin{eqnarray*}
D_k & := & 
\{
x \in B'_k : d (x, \partial B'_k) \ge 2 L_{k-1} 
\}
\\
B''_k &=& \{ 
x \in \Lambda_k : d(x, \partial (B'_k \setminus D_k)) \le L_{k-1} \},
\\
B'''_k &=& \{
x \in B'_k : d(x, \partial (B'_k \setminus D_k)) \le L_{k-1} \}.
\end{eqnarray*}
We take any 
$0 < \gamma' < \gamma$
and let
\begin{eqnarray*}
H'_k & := & H |_{B_k'}
\quad
\mbox{(periodic b.c.)}
\\
H''_{k}
& := & H |_{B''_k}, 
\quad
H'''_k := H'_k |_{B'''_k}
\quad
\mbox{(Dirichlet b.c.)}
\\
\epsilon_{k-1} & := & e^{- \gamma' L_{k-1}/2}.
\end{eqnarray*}
We first decompose
\begin{eqnarray}
N(H_k, J, B_k')
=
N(H_k, J, D_k)
+
N(H_k, J, B'_k \setminus D_k).
\label{decomposition}
\end{eqnarray}
To estimate the second term, 
we consider the following event
\begin{eqnarray}
\Omega'_k  & :=& 
\Biggl\{
\omega\in \Omega : 
\mbox{ For all $E \in I$, either }
\mbox{
$\Lambda_{L_{k-1}} (x)$
or
$\Lambda_{L_{k-1}}(y)$
is
$(\gamma, E)$-regular }
\nonumber
\\
&&
\mbox{any disjoint pair of boxes }
\Lambda_{k-1}(x), \Lambda_{k-1}(y)
\subset \Lambda_k \cup B'_k
\Biggr\}
\label{Omegakdash}
\end{eqnarray}
where 
$H_{\Lambda_{k-1}(x)}$, $H_{\Lambda_{k-1}(y)}$
have Dirichlet b.c.
As in (\ref{eventk}), we regard
$\Lambda_k, B'_k$ 
as torus. 
Since 
${\bf P}(\Omega_k^{'c}) \le (const.) L_{k-1}^{2 \alpha d - 2p}$, 
$p > 2d$,
$\omega\in \Omega'_0 := \liminf_{k \to \infty} \Omega'_k$ 
satisfies 
${\bf P}(\Omega'_0) = 1$, 
and for 
$\omega \in \Omega'_0$
we can find
$k'_0 (\omega)$
with
$\omega \in \Omega'_k$
if
$k \ge k'_0$. 
The following lemma
is proved similarly as 
Lemma \ref{many eigenvalues}. 
\begin{lemma}
\label{many eigenvalues4}
We can find 
$k_4(\alpha, d, \gamma, \gamma')$
such that, if
$k \ge k_4(\alpha, d, \gamma, \gamma')$
and
$\omega \in \Omega'_k$, 
we have
\begin{eqnarray*}
&(1)&\;
N(H_k, J,  D_k) \le N(H'_k, J+I(0, \epsilon_{k-1}))
\\
&(2)&\;
N(H_k, J, B'_k \setminus D_k) \le N(H''_k,  J+I(0, \epsilon_{k-1}))
\\
&(3)&\;
N(H'_k, J, B'_k \setminus D_k)
\le
N(H'''_k,  J+I(0, \epsilon_{k-1})).
\end{eqnarray*}
\end{lemma}
The following lemma is similar to Lemma \ref{many eigenvalues3} but additionally has a control on the location of  localization centers of the big box. 
\begin{lemma}
\label{many centers}
We can find 
$k_5(\alpha, d, \gamma, \gamma')$
such that, if
$k \ge k_5(\alpha, d, \gamma, \gamma')$
and 
$\omega \in \Omega'_k$, 
we have
\[
N(H'_k, J, D_k)
\le
N(H_k, J + I(0, \epsilon_{k-1}), B'_k).
\]
\end{lemma}
{\it Idea of proof of Lemma \ref{many centers}}\\
Let
\[
C_k := 
\{ x \in B'_k : d(x, \partial B'_k) \ge L_{k-1} \}, 
\]
let 
$M := N(H'_{k}, J, D_k)$
and let 
$P$
be the spectral projection of 
$H_{k}$
corresponding to 
$J + I(0, \epsilon_{k-1})$. 
Since
$\phi_1, \cdots, \phi_M \in {\cal E}f(H'_k, J, D_k)$
decay exponentially on 
$C_k^c$, so are 
$P \phi_1, \cdots, P \phi_M$.
We can write
\[
P \phi_1 = \psi_1 + \psi'_1, 
\cdots, 
P \phi_M = \psi_M + \psi'_M, 
\]
where
$\{ \psi_j \} 
\subset 
\mbox{Span }
{\cal E}f(H_{k}, J + I(0, \epsilon_{k-1}), B'_k), 
\{ \psi'_j \}
\subset 
\mbox{Span }
{\cal E}f(H_{k}, J + I(0, \epsilon_{k-1}), (B'_k)^c)$.
Since 
$\{ P \phi_l \}$
are ONS on 
$l^2 (C_k)$
modulo exponential error, 
and since 
$\psi'_j$
decays exponentially on 
$C_k$, 
$\psi_1, \cdots, \psi_M$
are linearly independent so that
$N(H_{k}, J+I(0, \epsilon_{k-1}), B'_k) \ge M$.
\QED
\\
We further take
$k \ge k_4(\alpha, d, \gamma, \gamma')
\vee
k_5(\alpha, d, \gamma, \gamma')$.
Since by Lemma \ref{many eigenvalues4}(2), 
$
N(H_k, J, B'_k \setminus D_k)
\le
N(H''_k, J+I(0, \epsilon_{k-1}))
\le
(const.) L_{k-1} L_k^{d-1}$, 
we have
\begin{equation}
\frac {1}{| \Lambda_k |}
N(H_{k}, J, B_k' \setminus D_k)
\le
\frac {L_{k-1}}{L_k} = o(1).
\label{error estimate}
\end{equation}
In what follows, 
we assume that the origin is the lower-left endpoint of 
$B$. 
By (\ref{IDS}) and by
Lemma \ref{many eigenvalues4}(1) it follows that, 
for any 
$\epsilon > 0$
\[
N(H_k, J, D_k)
\le
N(H'_k, J+I(0, \epsilon_{k-1}))
\le
|B| | \Lambda_k | (\nu (J) + \epsilon)
\]
for large
$k$. 
Together with
(\ref{equality}), (\ref{decomposition})
and
(\ref{error estimate}), 
we have
\begin{equation}
\limsup_{k \to \infty}
\frac {1}{| \Lambda_k |}
\bar{\xi}_k (J \times B)
\le
|B| \nu (J).
\label{upper bound}
\end{equation}
On the other hand, 
by Lemma \ref{many eigenvalues4}, \ref{many centers}, 
\begin{eqnarray*}
N(H'_k, J+I(0, \epsilon_{k-1}))
&=&
N(H'_k, J+ I(0, \epsilon_{k-1}), D_k)
+
N(H'_k, J+I(0, \epsilon_{k-1}), B'_k \setminus D_k)
\\
& \le &
N (H_k, J+I(0, 2\epsilon_{k-1}), B'_k)
+
N(H'''_k, J+I(0, 2\epsilon_{k-1}))
\\
& \le &
N(H_k, J+I(0, 2\epsilon_{k-1}), B'_k)
+
(const.) L_{k-1} L_k^{d-1}.
\end{eqnarray*}
Hence for large
$k$, 
\[
N(H'_k, J - I(0, \epsilon) )
\le
N(H_k, J, B'_k)
+
(const.) L_{k-1} L_k^{d-1}
\]
for any 
$\epsilon>0$. 
Dividing by
$| \Lambda_k |$
and letting 
$k \to \infty$, 
we have
\begin{equation}
|B|( \nu (J) - \epsilon)
\le
\liminf_{k \to \infty} 
\frac {1}{| \Lambda_k |}
\bar{\xi}_k (J \times B).
\label{lower bound}
\end{equation}
(\ref{upper bound}), (\ref{lower bound})
prove 
(\ref{suffice})
if the origin is the lower-left endpoint of 
$B$. 
For general
$B$, 
(\ref{suffice})
follows from a subtraction argument.\\

\noindent {\bf Acknowledgement }
The author 
would like to thank professors Rowan Killip, Nariyuki Minami and a referee for helpful discussions and comments. 
This work is partially supported by 
JSPS grant Kiban-C no.18540125.
%

\end{document}